\documentclass[a4paper]{jpconf}
\usepackage{graphicx}
\begin{document}
\title{Interpretation of surface diffusion data with Langevin simulations
- a quantitative assessment}




\author{\textbf{M. Diamant$^1$, S. Rahav$^1$, R. Ferrando$^2$ and G. Alexandrowicz$^{1+}$}}

\address{\textbf{$^1$Schulich Faculty of Chemistry, Technion - Israel Institute of Technology, Technion City, Haifa 32000, Israel.}}

\address{\textbf{ $^2$Dipartimento di Fisica, Universita di Genova - v. Dodecaneso 33 - 16146, Genova, Italy .}}

\ead{$^+$ga232@tx.technion.ac.il}

\begin{abstract}
Diffusion studies of adsorbates moving on a surface are often analyzed
using 2D Langevin simulations. These simulations are computationally cheap and offer valuable insight into the dynamics, however, they simplify the complex
interactions between the substrate and adsorbate atoms, neglecting correlations in the motion of the two species. The effect of this simplification on the accuracy of observables extracted using Langevin simulations was previously unquantified.
Here we report a numerical study aimed at assessing the validity of
this approach. We compared experimentally accessible
observables which were calculated using a Langevin simulation with
those obtained from explicit molecular dynamics simulations. Our results
show that within the range of parameters we explored Langevin simulations provide a good alternative for calculating the diffusion procress, i.e. the effect
of correlations is too small to be observed within the numerical accuracy
of this study and most likely would not have a significant effect on
the interpretation of experimental data. Our comparison of the two
numerical approaches also demonstrates the effect temperature dependent
friction has on the calculated observables, illustrating the importance
of accounting for such a temperature dependence when interpreting
experimental data.
\end{abstract}

\section{Introduction and motivation}

The diffusion of atoms and molecules on surfaces is of importance in a wide range of research fields and applications, consequently, a wide range of dedicated experimental and theoretical methods have been developed over the years
\cite{Antczak_book,ala-nissila_collective_2002}. One of these techniques is quasi elastic helium atom scattering (QHAS). This method which has received a significant boost with the availability
of the helium spin echo (HSE) apparatus\cite{alexandrowicz_helium_2007,jardine_studying_2009,jardine_helium-3_????},
provides a unique opportunity to follow atomic scale
motions on time scales of pico to nano-seconds.
With new data comes the need for new or improved models to interpret
the data and extract the underlying physical properties of the surface
system. One commonly used interpretation model is based on a 2D Langevin
simulation which allows the extraction of a potential energy surface
term and a friction parameter from the experimental data.
One obvious drawback of this model is that the complex dynamics of
the substrate atoms are not explicitly treated and correlations between
the motion of the adsorbate and substrate particles can not be accounted
for. Another approach, which to the best of our knowledge was only
applied twice to interpret QHAS measurements, is using molecular dynamics
(MD) simulations\cite{ellis_molecular_1994,Fouquet_2009}. In these MD simulations, the motion of the surface
atoms and their interaction with their neighbors are explicitly calculated
and correlation effects between the motion of the adsorbate and the
substrate atoms are inherently included.
On the down side, these explicit MD simulations are computationally
expensive. In this manuscript we describe a numerical study which
aims to quantify the differences between these two approaches and
probe the validity of using the simpler and less time consuming Langevin
approach. This comparison is performed by calculating the observables
with both simulations under similar conditions. The paper is organized
in the following manner, we start by introducing some useful relations
and definitions, then we describe the two numerical simulations and
explain how they were tuned to simulate similar surface systems and
finally we present the results of the comparison.

\section{\label{sec:Basic-definitions-and}Basic definitions and methods for
spectra interpretation.}

Surface diffusion is a general process which describes the motion
of particles ranging from atoms to macro molecules which are confined
to a surface\cite{Antczak_book,ala-nissila_collective_2002}.
For most systems surface diffusion is essentially a classical process
- with Hydrogen diffusion at low temperatures being an example of
an exception \cite{jardine_determination_2010}. For molecular adsorbates
a simple and sometimes sufficiently good description can be obtained
by ignoring the internal degrees of freedom, although it should be
noted that these degrees of freedom can play an important role in some systems \cite{backus_real-time_2005}.

There are several physical properties which are accessible to experiments
and can be used to characterize surface motion, particularly popular
choices are the the tracer diffusion coefficient in the case of isolated
diffusion and the chemical diffusion coefficient for the case of collective
motion. Another way of characterizing motion is using pair correlation
functions which, unlike the diffusion coefficients mentioned above,
contain a full statistical description of the motion and its underlying
mechanism. The pair correlation function we use can be written as
a sum of the self correlation and the distinct correlation functions,
$G(\bar{R},t)=G{\scriptscriptstyle s}(\bar{R},t)+G{\scriptscriptstyle d}(\bar{R},t)$
defined by \cite{van_hove_correlations_1954,hulpke1992helium}{\small{}
\begin{eqnarray}
G_{s}\left(\vec{R},t\right) & = & \frac{1}{N}\left\langle \sum_{i=1}^{N}\delta\left(\vec{R}+\vec{R}_{i}\left(0\right)-\vec{R}_{i}\left(t\right)\right)\right\rangle \label{eq:van hove}\\
G_{d}\left(\vec{R},t\right) & = & \frac{1}{N}\left\langle \sum_{i=1}^{N}\sum_{j\neq i}^{N}\delta\left(\vec{R}+\vec{R}_{i}\left(0\right)-\vec{R}_{j}\left(t\right)\right)\right\rangle 
\end{eqnarray}
}{\small \par}

These correlation functions can be interpreted as a measure of the probability of finding a particle at location $\vec{R}$ at
time $t$, given that the same $\left(G_{s}\right)$ or a different
$\left(G_{d}\right)$ particle was at the origin at time $t=0$ .
The self correlation function describes the complete dynamics of an
individual adsorbate. It is also the dominant contribution for dilute
adsorbate coverages. In this work we will focus on the zero coverage
limit, neglecting the contribution from the distinct correlation function.
One advantage of using these pair correlation functions is their close
relation to quantities which can be measured in experiments. In particular
Fourier transforming $G_{s}(\bar{R},t)$ to the momentum domain gives
$I_{s}\left(\Delta\vec{K},t\right)$, the self part of the intermediate
scattering function (ISF) {\small{}
\begin{eqnarray}
I_{s}\left(\Delta\vec{K},t\right) & = & \frac{1}{N}\left\langle \sum_{i=1}^{N}e^{-i\Delta\vec{K}\cdot\vec{R}_{i}\left(0\right)}e^{-i\Delta\vec{K}\cdot\vec{R}_{i}\left(t\right)}\right\rangle \label{eq:ISF self}
\end{eqnarray}
}where $\Delta\vec{K}$ is proportional to the momentum exchange parallel
to the surface in a scattering experiment. Within the kinematic scattering
approximation it can be shown that the HSE technique mentioned above
measures this quantity directly\cite{alexandrowicz_helium_2007,jardine_studying_2009,jardine_helium-3_????},
where the experimentally accessible $\Delta\vec{K}$ values range
between 0.01 to a few inverse angstroms and the times, $t$, range
from 0.1ps to a few nano seconds. Using the same scattering approximation
it can be shown that for a time of flight Helium atom scattering apparatus,
the observable quantity is the temporal Fourier transform of $I\left(\Delta\vec{K},t\right)$
known as the Dynamic Structure Factor (DSF), $S\left(\Delta\vec{K},\omega\right)$,{\small{}
}where $\hbar\omega$ is the energy exchange during the scattering
event. 

Generally speaking, when random motion takes place on a surface
$I\left(\Delta\vec{K},t\right)$ decays with time reflecting the loss
of correlation of the position of the surface particles, a decaying
ISF corresponds to a peak in the DSF which is centered around $\omega=0$
and has a width which is inversely proportional to the decay rate.
This peak is called the quasi-elastic peak (QEP) and its width, $\Gamma$,
is often termed the quasi-elastic broadening. Whether one measures the ISF (using HSE \cite{alexandrowicz_helium_2007,jardine_studying_2009,jardine_helium-3_????})
or the DSF (using time of flight helium scattering \cite{Graham2003115}),
an analytical or numerical model is needed to extract the physical
properties of the surface dynamics from the data.

One particularly useful analytical model for surface diffusion of
adsorbates, which can be used to calculate the quasi-elastic broadening, is the jump diffusion model which was first derived by
Chudley and Elliott \cite{chudley_neutron_1961} for neutron scattering
measurements of bulk diffusion. In this model vibrational motions
within the adsorption sites (intra-cell motions) are ignored and the
inter-cell jumps between one adsorption site and another are assumed
to be instantaneous. Generally speaking, the Chudley Elliot model
is suited for systems where i) the energy barrier for diffusion is
large compared with the thermal energy ii) the adsorption sites form
a Bravais lattice and iii) the adsorbate coverage is sufficiently
small to not be affected by the presence of other adsorbates, i.e.
we can ignore the influence of the distinct pair correlation function.
The results of this simple model are an ISF which decays exponentially
with time, $I_{s}\left(\Delta\vec{K},t\right)\propto e^{-\Gamma\left(\Delta K\right)t}$
which corresponds to a DSF which contains a Lorentzian peak centered
at $\omega=0$ with a finite width, $\Gamma\left(\Delta K\right)$
. For the case of the Chudley-Elliot model the dependence of the quasi
elastic broadening on the momentum transfer is given by

\begin{equation}
\Gamma\left(\Delta K\right)=4\hbar\sum_{j}\nu_{j}\sin^{2}\left(\frac{\Delta\vec{K}\cdot\vec{j}}{2}\right)\label{eq:Jump QHAS-1}
\end{equation}

{\small{}where }the sum is over of a discrete set of $j$ possible
jump vectors connecting two adsorption sites and $\nu{}_{j}$ is the
jump rate for a particular jump vector. Eq. (\ref{eq:Jump QHAS-1})
contains all the information about the jump diffusion process, hence,
the experimental ISF allows us to extract the different jump rates%
\footnote{Generally we need to measure along two different crystal azimuths
if we wish to calculate the jump rates of the various jump proceses.%
}. Furthermore, if we make the assumption that the jump rate $\nu_{j}$
is of the form $Ae^{-\frac{k_{B}T}{E_{b}}}$ we can find the potential
barrier $E_{b}$ the adsorbate has to overcome by finding the temperature
dependence of the $\nu_{j}$ coefficients.

\section{Numerical models and interpretation methods}

An analytic approach in general and eq. (\ref{eq:Jump QHAS-1}) in
particular, provides significant insights into the dynamics when analyzing
experimental data (e.g. \cite{alexandrowicz_observation_2004}). On the
other hand using equation \ref{eq:Jump QHAS-1}, which treats the
surface as a discrete set of point-like adsorption sites that an isolated
point-like particle jumps between, is typically restricted to relatively
simple surface dynamics systems where this ideal-jump model is valid.
As mentioned above, an alternative approach which has been extensively
used in the last few decades is to numerically calculate the trajectories
using a set of parameters which describe the various interactions,
extract observables such as the DSF or ISF from the trajectories and
by comparison with the experimental observables improve the interaction
parameters until a good fit is obtained. For practical reasons the
comparison is typically performed on 1D quantities such as the dependence
of the quasi-elastic broadening as function of momentum transfer,
temperature or coverage rather than direct comparison of the two dimensional
DSF or ISF functions\cite{jardine_studying_2009,alexandrowicz_helium_2007}.

The two numerical approaches we compare in this work are molecular
dynamics (MD) and 2D Langevin simulations. The first, provides an
explicit treatment of the interactions between all the particles,
whereas the second provides fast computation and a simple separation
of the static and dynamic interactions characterizing the surface
and has been heavily used to interpret quasi-elastic helium scattering
experiments \cite{alexandrowicz_helium_2007,jardine_studying_2009,jardine_helium-3_????}.

\subsection{The molecular dynamics model}

MD simulations include the degrees of freedom of both the adsorbate
and substrate atoms. In order to mimic experimental systems in a realistic
way, complex many-body interactions can be used \cite{doi:10.1080/01418618908205062}.
However, the purpose of this work is to study how well Langevin simulations
can reproduce the explicit approach of MD modeling. This can be studied
with particularly simple interactions for the MD simulation; pair-wise
harmonic interaction between the substrate atoms and a Morse potential
between the adsorbate and the substrate atoms. The parameters of these
two interaction models and the mass of the particles were initially
chosen to resemble an experimentally relevant situation - the motion
of a sodium atom (mass = 23 amu) on a flat (001) copper surface. The
Na/Cu(001) system has been extensively studied experimentally, both
in the regime of low coverage where the sodium atom can be assumed
to move as an isolated adsorbate\cite{ellis_observation_1993,graham_experimental_1997}
and at higher coverages where correlated motion effects dominate \cite{alexandrowicz_onset_2006}.
Furthermore, this system represents a rare case 
where a MD simulation has been used to interpret quasi-elastic helium
scattering measurements, which conveniently supplies us with a set
of parameters for both the harmonic and the Morse potentials \cite{ellis_molecular_1994,ellis_observation_1993}. 

For the harmonic term, $0.5k(r-r_0)^2$, a single force constant between nearest neighbohrs of $k=28\frac{N}{m}$
was shown to provide a good description of the copper bulk phonons
\cite{white_atomic_1958,ellis_molecular_1994,svensson_crystal_1967}.
As mentioned above, the surface-adsorbate potential was modeled with
a Morse like potential 
\begin{eqnarray}
V\left(r\right) & = & \sum_{j}A\left(e^{-2\beta\left(r_{j}-r_{0}\right)}-2e^{-\beta\left(r_{j}-r_{0}\right)}\right)\label{eq: Morse potential}
\end{eqnarray}
where $r_{j}$ is the distance between the j'th substrate atoms and
the adsorbate and the sum runs over all the substrate atoms. The following
values were used for the parameters\cite{ellis_molecular_1994}
\begin{eqnarray}
A & = & 0.135eV\nonumber \\
\beta & = & 0.875\AA^{-1}\label{eq:Morse parameters}\\
r_{0} & = & 3.3\AA\nonumber 
\end{eqnarray}
which reproduce the experimental measurements of the adsorption height
and vibrational frequencies . 

The geometry we used included a copper solid consisting of 7 layers
of $8\times8$ lattice cells with a total of 896 atoms. Periodic boundary
conditions were imposed parallel to the surface. The bottom layer
of the slab was frozen to simulate the rest of bulk layers and to
fix the center of mass in its place. The substrate atoms at $t=0$
were placed in their bulk equilibrium positions, and were given a
random initial velocity using a Maxwell-Boltzmann distribution. The
atoms were then allowed to relax, after this relaxation period a single adsorbate atom was added on
the surface and the system was allowed to relax again to the desired
temperature. The simulation was carried in the micro canonical ensemble.
The Newtonian equations of motion were solved using Beeman's algorithm
\cite{frenkel2001understanding} for both substrate atoms and the
single adsorbate.

\subsection{The Langevin model}

A popular approach for interpreting quasi-elastic helium scattering
experiments is using a 2D Langevin simulation. When inter-adsorbate
interactions can be ignored (the zero coverage limit), the force is
given by
\begin{eqnarray}
\dot{\vec{p}}\left(t\right) & = & -\vec{\nabla}\phi\left(\vec{r}\right)-\eta\vec{p}\left(t\right)+\vec{f}\left(t\right).\label{eq:Langevin eom}
\end{eqnarray}
where $\vec{p}\left(x,y\right)$ is the adsorbate's 2D momentum. The
equation includes a constant potential energy surface (PES) term $\phi\left(x,y\right)$
which is the potential the adsorbate experiences when the surface
atoms are at their equilibrium points (See sec. (\ref{sec:Constructing-a-comparable})
for a description of the procedure used to obtain the PES). The two
terms which replace the explicit treatment of the dynamic interaction
between the adsorbate and substrate atoms are a dissipation term $\eta$
which leads to energy losses and a random fluctuating force $\vec{f}$
(typically chosen as a white noise force) which allows energy to be
supplied to the adsorbate. These two terms are not independent, as
they are related through the fluctuation dissipation theorem\cite{0034-4885-29-1-306}
\begin{eqnarray}
\left\langle f_{i}\left(t\right)f_{j}\left(t'\right)\right\rangle  & = & 2\eta k_{B}T\delta_{i,j}\delta\left(t-t'\right).\label{eq:fluc diss relation}
\end{eqnarray}
It should be noted that if the issue of restricted computational time
is ignored it would be preferable to extend the simulation to include
a three dimensional motion of the adsorbate. However, since the height
of an adsorbate above the surface is typically restricted to a very
narrow range (fractions of an Angstrom) and correspondingly the vibrational
period perpendicular to the surface is about an order of magnitude
faster than the motion of the adsorbate parallel to the surface, the
effect of the vertical motion is typically assumed to be averaged
out and a 2D Langevin approach is used for analysis. As this is the
most frequently used approach we chose to use it in our comparison.

When fitting experimental data with a Langevin simulation (in the
zero coverage limit), the free parameters are those used to define
the PES and the friction parameter $\eta$. The parameters of the
PES provide important insight into the average interaction between
an adsorbate and the surface, they can be used to compare the energy
of multiple adsorption sites \cite{alexandrowicz_observation_2008}
and provide an important benchmark for density functional theory calculations
\cite{PhysRevB.80.045422}. The friction parameter reflects the atomic-scale
energy transfer mechanism and plays an important role in a wide range
of research fields and applications\cite{Krim_review}. Since measurements of atomic scale friction of isloated adsorbates are scarse, the ability to extract such values from Langevin analysis of quasi-elastic
helium scattering measurements is particularly important\cite{hedgeland_measurement_2009}.

The current theoretical understanding of surface friction is rather
limited, it is however custom to separate the frictional coupling
into two main contributions, namely, electronic and phononic friction.
Within the Langevin approach, the friction coupling is a fitting parameter
and its value reflects the total friction regardless of its origin,
this is in contrast with MD simulations where the friction is not
a parameter, rather it is a result of the explicit interactions between
atoms. Consequently, since typically the interactions calculated in
MD simulations are between the ions, the only friction mechanism which
is simulated is phononic friction and systems where electronic friction
is important can not be accurately studied with simple MD simulations
of the type described above.

\subsubsection{Co\label{sec:Constructing-a-comparable}nstructing a comparable Langevin
simulation}

As mentioned above, in a Langevin simulation the energy transfer due
to the substrate motion or other mechanisms is accounted for using
a damping term and a fluctuating force term, both of which are determined
by a single friction parameter, $\eta$. In this work, the friction
parameter is used as an adjustable fitting parameter. The Langevin
simulation also requires the adiabatic interaction potential, i.e.
the PES. When analyzing experimental data, the PES is derived using
adjustable fitting parameters, however in this study, our goal is
to perform a relevant comparison with a specific MD model. In order
to do this we chose the PES to be the time averaged potential in the
MD simulation. The procedure for deriving the PES is the following:
\begin{enumerate}
\item Generate a 2D grid above the periodic unit of the substrate top layer.
\item For every point in the grid, fix the lateral coordinates of the adsorbate,
leaving all other degrees of freedom free. Allow the system to relax
to the equilibrium geometry. %
\footnote{The relaxation is performed by calculating the product $F_{ij}V_{ij}$
in the explicit MD simulation, where $F_{ij}$is the force acting
on atom $i$ along its free coordinate $j$ in the system and $V_{ij}$
is the velocity. If this product is positive then the atoms are moved
according to the force, if it is negative the velocity is set to zero.
The quenching procedure described in the previous stage continues
until the change in the system's total energy between time steps drops
to a negligible level. The forces acting on the atoms are the same
forces used in the explicit MD simulation described above.%
} 
\item The value of the PES at the grid point is set to be the potential
energy of the entire system - contributions from the adsorbate-substrate
interaction as well as interaction between the substrate's atoms. 
\end{enumerate}
Using this procedure and the interaction parameters mentioned earlier
(for the harmonic and Morse interactions) the potential difference
between a local minimum and saddle point of the PES was found to be
75 meV\footnote{Using the dimensionless quantity, $E/(k_B T)$, this energy barrier corresponds to values in the range 2.9-6.2 for the temperatures this study was performed at (from 300K down to 140K)}  . The same PES was used for all adsorbate masses in this work.

\subsection{Interpretation method}

\begin{figure}[h]
\includegraphics[width=17.2cm]{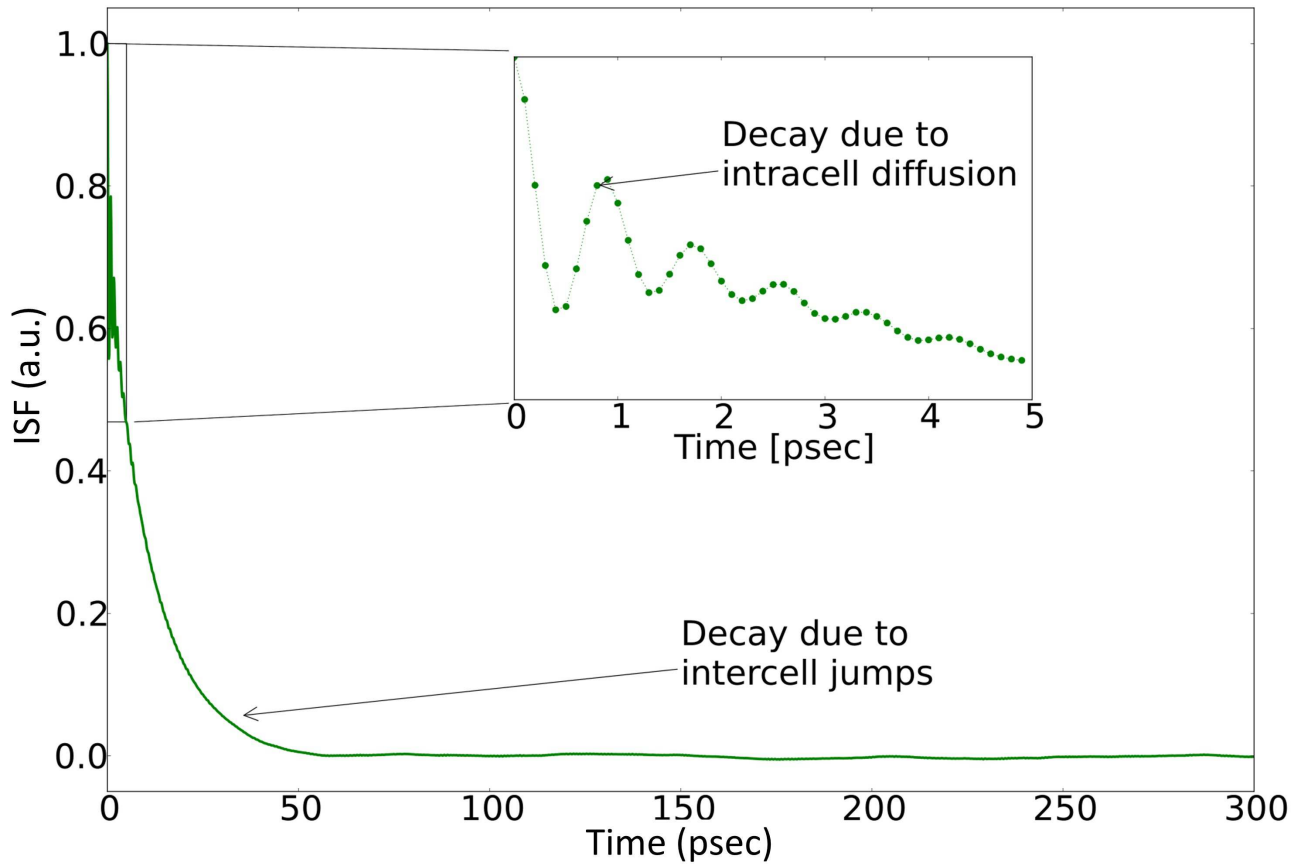}\\

\protect\caption{\label{fig:ISF}Example of an ISF function. The inset shows the ISF
at short times, where a combination of decaying oscillations as well
as a decaying exponential can be seen. Their origin is explained in
the text.}
\end{figure}

Both of the simulations used in this work generate trajectories of
the adsorbate. From each trajectory we can construct the ISF of the
adsorbate using eq. (\ref{eq:ISF self}). Figure (\ref{fig:ISF})
shows an example of the ISF from a 10 nanosecond trajectory calculated
by the Langevin simulation. The ISF contains 2 main features with
different characteristic time scales i) a slow decay of the ISF which
takes place over tens of pico seconds and ii) a rapid initial decay
and an oscillatory pattern, which can be seen more clearly in the
inset in figure (\ref{fig:ISF}) which depicts the ISF at short times.
Both of the features mentioned above are characteristic of surface
diffusion systems and have been seen in experimental and theoretical
work \cite{propane_NJP,jardine:8724,jardine_studying_2009}.

\begin{figure}[h]
\begin{centering}
\includegraphics[width=17.2cm]{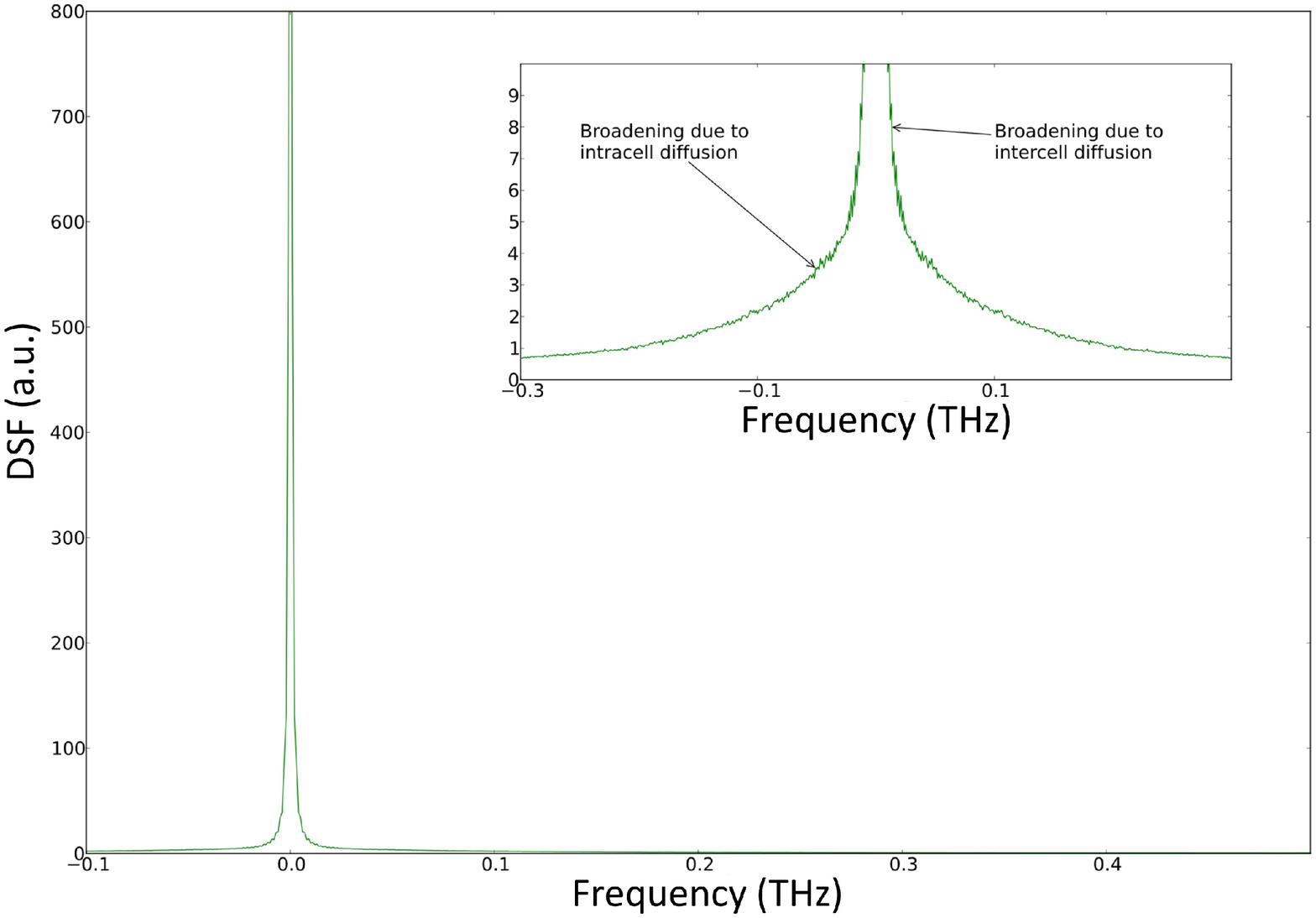}
\par\end{centering}

\protect\caption{\label{fig:DSF}Example of a DSF function calculated from the ISF
shown in figure \ref{fig:ISF}, focusing on the two peaks which are
located at the origin. The much sharper QEP dominates the DSF, whereas
the underlying broad QEB can be seen more clearly in the inset plot.}
\end{figure}

The slow exponential decay is due to intercell diffusion i.e. transitions
between local minimum in a corrugated potential as was discussed in
section \ref{sec:Basic-definitions-and}. The quasi-elastic broadening,
(or the decay rate of the ISF), $\Gamma$, and its dependence on $\Delta K$
and T can be related to the dynamics either using simple analytical
theory (e.g. equation \ref{eq:Jump QHAS-1}) or as will be demonstrated
below using more detailed numerical models. In the following section
we will use $\Gamma\left(\Delta K,T\right)$ to compare the diffusion
process calculated by the two numerical simulations. 

The oscillation and decay seen at short time scales, is related to
the motion within the adsorption site. The oscillation period reflects
the vibrational motion of the adsorbate whereas the fast decay reflects
the loss of phase coherency due to the random nature of this motion,
a process sometimes referred to as intra-cell diffusion \cite{jardine:8724,propane_NJP}.
In the DSF this intra-cell motion appears as three peaks, two inelastic
peaks located at the energy gain / loss values which correspond to
vibrational frequency and one additional peak centered at $\omega=0$,
as shown in figure (\ref{fig:DSF}). The width of all three peaks is
related to the rapid decay due to the phase loss of the intra-cell
motion mentioned above. Since this decay is typically much faster
than that due to the inter-cell motion, the widths of all three peaks
are substantially larger than that of the QEP (i.e. the quasi-elastic
broadening). In this work we will refer to the intra-cell motion contribution
centered at $\omega=0$ as the quasi-elastic base (QEB) to differentiate
it from the much sharper QEP which is also centered at $\omega=0$.
As mentioned above, in many cases the time scale of the intercell
diffusion is much slower compared to the intra-cell one, and the differentiation
of the different contributions mentioned above is valid%
\footnote{An exception to this case is when the temperature is sufficiently
high that the adsorbate's thermal energy is comparable with the corrugation
of the potential.%
}. In section \ref{sub:Estimating-the-friction} we will make use of
this separation scheme in order to extract values for the frictional
coupling within the adsorption site.

\section{Comparison of MD and Langevin quasi-elastic broadenings \label{sub:Comparison-of-the inter cell jumps}}

As mentioned above, under many circumstances, including the conditions
encountered in this work, inter-cell motion leads to an exponentially
decaying ISF equivalent to a Lorentzian QEP in the DSF. Under these
conditions, the quasi-elastic broadening $\Gamma$, which can be extracted
either from the decay rate of the ISF or from the width of the QEP
peak in the DSF,  can be used to characterize the inter-cell motion
from both experiments and theory \cite{alexandrowicz_helium_2007,jardine_studying_2009,jardine_helium-3_????}.
A method which allowed us to reliably extract , $\Gamma$, from the
calculated ISF is to delay the fitting procedure to times which are
sufficiently long to avoid mixing the contributions of the intra-cell
motion mentioned above. %
\footnote{In practice, the ISF at the time interval $[t_{0},t_{final}]$ was
fitted to a single exponential, with $t_{0}$ being advanced in time
at each iteration until the decay times between successive iterations
differed by less than 1\%%
}.

\begin{figure}[h]
\noindent \begin{centering}
\includegraphics[width=17.2cm]{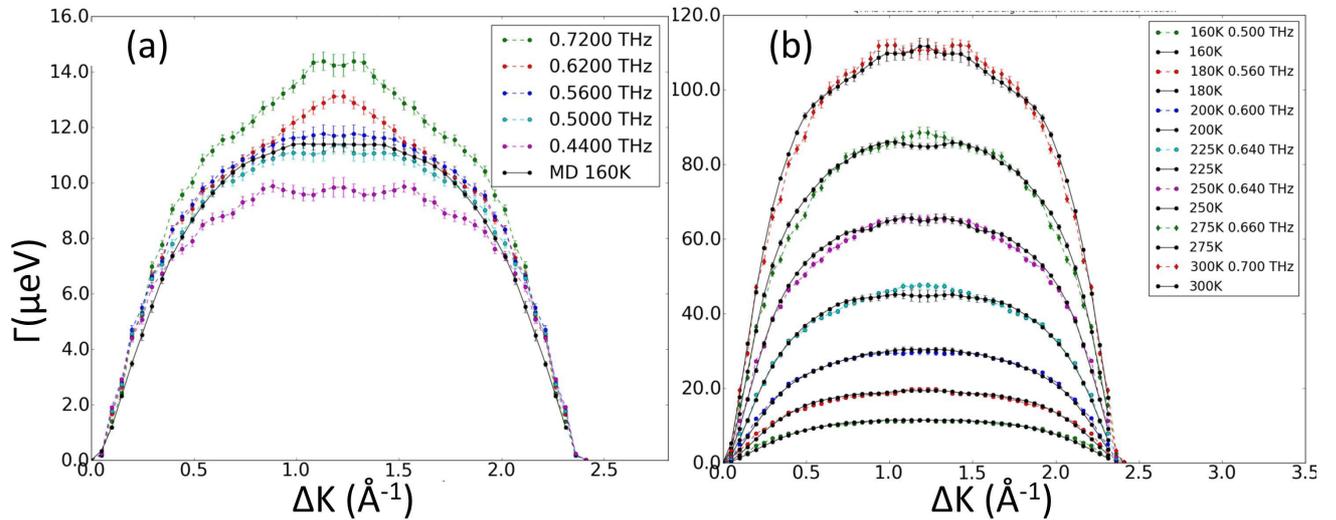}
\par\end{centering}

\protect\caption{\label{fig:Comparison optimal friction 23 amu} a) Quasi-elastic broadening,
$\Gamma\left(\Delta K\right)$, calculated along the (1,1,0) crystal
azimuth for a 23 amu adsorbate at 160K. MD results are shown with
full black circles alongside with the results of Langevin simulations
with different friction parameters as indicated in the legend. b)
Comparison between the $\Gamma\left(\Delta K\right)$ calculated by
the MD (black circles) and those calculated by the Langevin using the optimal friction
values indicated in the legend alongside the relevant surface temperature. }
\end{figure}

We start with the case of an adsorbate with a mass of 23 amu, representing
the Na/Cu(001) system mentioned earlier. Figure \ref{fig:Comparison optimal friction 23 amu}
shows a comparison of $\Gamma$ calculated using the two simulation
approaches. The left panel shows an example for calculations performed
at 160K, the MD results are shown using the black dot symbols, where
as Langevin results using different friction values in the range 0.44THz-0.72THz
are plotted with coloured symbols according to the legend. One immediate
feature which can be seen for both simulations is the oscillatory
nature of the width of the QEP as function of the momentum transfer
value. This is a characteristic feature of jump diffusion as can be
seen from the Chudley Elliot equation (\ref{eq:Jump QHAS-1}). A second
observation which can be made is that the Langevin simulation can
reproduce the MD result quite well if the friction parameter is set
to a value of 0.56THz \footnote{which for an oscillation frequency of $\nu=1.2THz$, can be expressed with the dimensionless quantity  $\eta / \nu = 0.47$. }  , we will refer to the friction parameter which
provides the best fit as the ``optimal friction value'', $\eta_{opt}$
. This particular value  is consistent with the results obtained in
the past when analyzing experimental measurements of Na/Cu(001) with
Langevin simulations \cite{alexandrowicz_onset_2006,graham_experimental_1997}.
For lower friction values we observe a slower jump rate due to weak
coupling between the substrate and adsorbate, while for higher friction
values the shape of the curve is narrower, indicating the dominance
of single jumps (equation (\ref{eq:Jump QHAS-1}) reverts to a single
oscillating term when only nearest neighbor jumps take place). 

The right panel of figure \ref{fig:Comparison optimal friction 23 amu} shows the same comparison in the temperature range
140K-300K. For each temperature we plot the MD calculation together
with the Langevin simulation obtained using the optimal friction values, $\eta_{opt}(T)$, i.e. the friction values which
gave the minimal standard deviation between the two $\Gamma\left(\Delta K\right)$
curves. Again one can see
that the Langevin simulation can reproduce the MD values quite well,
however, the optimal friction parameters (indicated in the legend)
are not identical for the different temperatures, instead there is
a subtle but clear trend where the friction parameter, $\eta_{opt}$,
increases with the temperature, i.e. the Langevin simulation we used
can not exactly reproduce the MD results if a single temperature independent
friction value is used.

\subsection{Temperature dependent friction }

\begin{figure}[h]
\begin{centering}
\includegraphics[width=10cm]{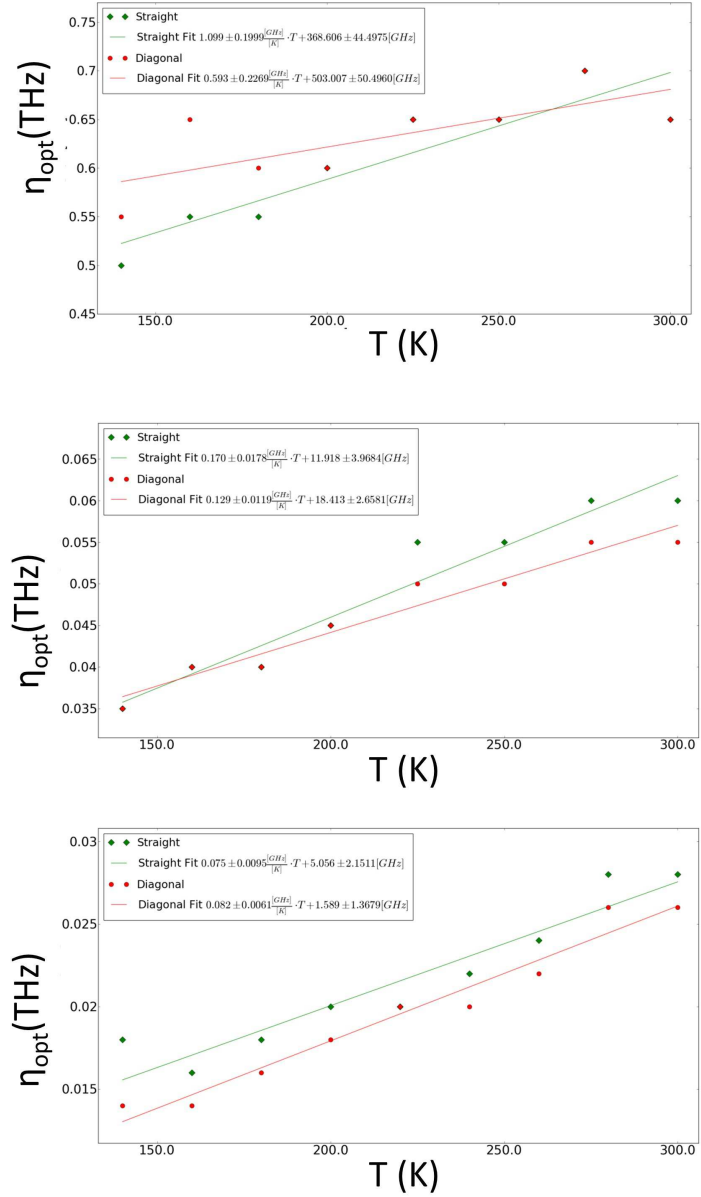}
\par\end{centering}

\protect\caption{\label{fig:Optimal friction vs temperature all masses}From top to
bottom: The optimal friction as a function of temperature for 23,
100 and 200 amu adsorbates respectively. Results for the "straight" (over the bridge site) crystal
azimuth (110) are plotted in green and results for the "diagonal" (over the top) azimuth (100) in red. A linear
form was fitted for both crystal azimuths. Note that the relative
change of the optimal friction parameter in the temperature range
150-300K can be as large as 100\% for the heavier adsorbates we simulated. }
\end{figure}

In the previous section we saw that we can find a good agreement between
the quasi-elastic broadenings, $\Gamma\left(\Delta K,T\right)$, calculated
by the two numerical models with only one free parameter, $\eta$
- the frictional coupling. However, in order to optimize the fit we
had to slightly adjust the friction according to the temperature.
During the last two decades various systems have been measured using
QHAS, most of which were analyzed using Langevin simulations where
a single, temperature independent, friction coefficient was assumed
\cite{alexandrowicz_helium_2007,jardine_studying_2009,jardine_helium-3_????}.
If the temperature dependence of the friction is significant for some
of these systems, the analysis method which was applied in the past
to extract an activation energy for these systems, resulted in a small
but systematic error which needs to be taken into account. In order
to study and understand this apparent temperature dependence we performed
further calculations for heavier adsorbates, as this allows us to
change the strength of the frictional coupling \cite{PhysRevB.32.3586,PhysRevE.65.061107}
while leaving the inter-atomic forces unchanged. Figure \ref{fig:Optimal friction vs temperature all masses}
shows the friction values which give the best quasi-elastic broadening
match between the two simulations at different temperatures for 100
amu and 200 amu adsorbates. The resolution of the friction parameter
is $5GHz$ and $2GHz$ for the 100 and 200 amu masses respectively.

Two Main observations can be made when comparing the results of the
different masses: i) The friction values needed to fit the two simulations
are significantly reduced for heavier adsorbates. This is the expected
trend, since heavier adsorbates have a lower vibration frequency and
are expected to have a weaker coupling to the substrate \cite{PhysRevB.32.3586,PhysRevE.65.061107}.
ii) The need to adjust the friction parameter according to the temperature
in order to get an agreement between the two simulations is more pronounced
for the heavier adsorbates. Thus, this temperature dependent friction
which is rather subtle for the 23 amu adsorbate, and would have a
small effect on the interpretation of experimental data , becomes
a more significant effect for heavier adsorbates.

\subsubsection{Estimating the friction from the MD simulation\label{sub:Estimating-the-friction}}

We have shown above, that in order to mimic the MD results using a
Langevin simulation we need to allow the friction parameter to increase
with temperature. One explanation for this is that by changing the
friction we simply make use of our only free parameter to compensate
for the fact that the Langevin simulation can not exactly mimic the
MD results, either due to the fundamental differences between the
two, or due to our particular choice for the PES. On the other hand,
since the friction is not an explicit parameter in the MD simulation,
another possibility is that the friction coupling changes with temperature
in the MD simulation and that the comparison with the Langevin simulation
is revealing this trend. In order to try and differentiate between
these possibilities we have attempted to extract an effective ``friction
parameter'' from the MD simulation and study its temperature dependence. 

We achieve this by extracting the width of the quasi-elastic base
(QEB), as mentioned in section (\ref{sec:Basic-definitions-and}).
The width of the QEB is governed by the dephasing rate of the intracell
motion, i.e. it is related to the friction coupling within the adsorption
site, a relation which has been shown, both analytically and numerically
\cite{propane_NJP,vega:8580}.
In fact, if one looks at the lowest order of the analytically derived
expression for the DSF, the half width of the QEB (in the angular frequency domain) is simply equal to $\eta$ \cite{vega:8580}. 

While the QEB width will undoubtedly be related to the frictional
coupling, the accuracy of the simple relation between the two mentioned
above is unknown. In particular, the analytical relation is valid
within certain approximations \cite{vega:8580}. Furthermore, the
friction in the Langevin simulation reflects the average energy exchange
rate, both within and outside the adsorption site, whereas the QEB
is only related to the intracell motion within the adsorption site, hence the two properties are obviously not identical.
In order to validate our approach, we first start by applying this
method on the DSF calculated by Langevin. Since Langevin simulations
include an explicit friction value, our ability to reproduce this
value from the QEB acts as a self consistency check for our method.

\begin{figure}[h]
\begin{centering}
\includegraphics[width=17.2cm]{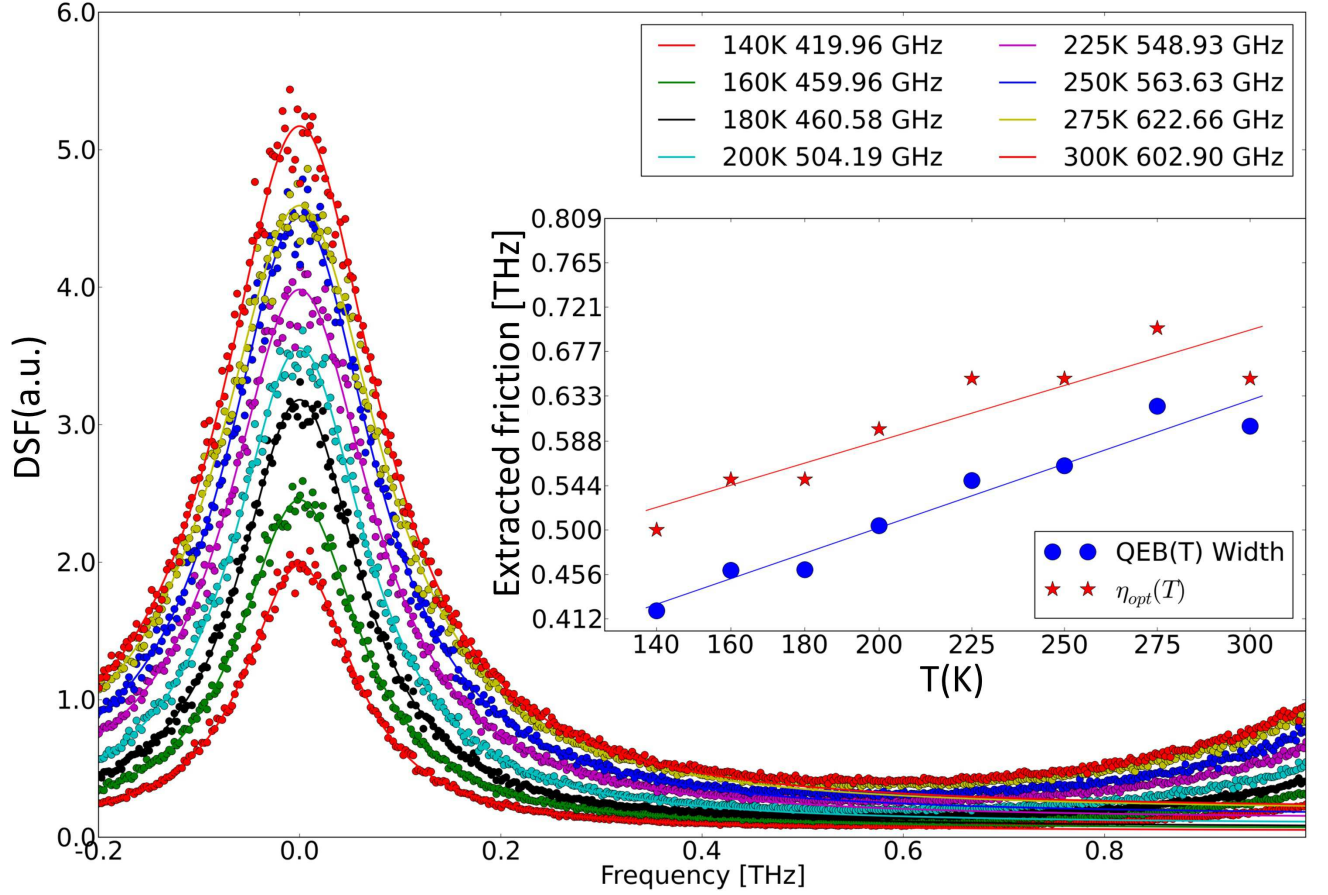}
\par\end{centering}

\protect\caption{\label{fig:Langevin QEB} Fitting the QEB peak calculated by the Langevin
simulations for the different temperatures. The full symbols are the DSF values calculated by the Langevin
simulations, performed using  $\eta_{opt}$ friction parameters. The solid lines show the Lorentzian fit described in the text which was used to extract the QEB width.  The inset compares $\eta_{opt}$ with the friction values extracted from the width of the Lorentzian
peak which best fitted the QEB\cite{vega:8580}. }
\end{figure}

In order to assist the fitting procedure and separate any contributions
from intercell diffusion, the DSF calculations were performed along
the straight azimuth for $\Delta K\approx2.46\AA$, conditions under
which the QEP has a negligible width due to the jump diffusion process
(minima values of $\Gamma$ in eq. \ref{eq:Jump QHAS-1}). %
\footnote{The fitting range started at $d\omega$, where $d\omega$ is the frequency
resolution of the calculated DSF. This range was chosen to eliminate the QEP contribution
which manifests itself in the DSF as a single data point at $\omega=0$.
The fitting range extended to a frequency which provided enough data
points for the fit, yet avoided contribution from the Lorentzian centered
about the vibration frequency}. The fit for the Langevin data is shown in figure (\ref{fig:Langevin QEB}).
At each temperature, the DSF corresponds to a calculation using the
optimal friction value from figure (\ref{fig:Optimal friction vs temperature all masses}).
The inset in figure (\ref{fig:Langevin QEB}) shows the friction values
extracted from the QEB width (blue circles) versus the friction parameters
used in the simulation (denoted $\eta_{opt})$ . Overall, the two
values are very close, with the QEB underestimating the friction parameters
by 15\%-20\%, similar calculation for higher masses (100 amu and 200
amu, data not shown) produce even smaller deviations between the two.
Thus, we conclude that the QEB width provides a reasonable way to
estimate the friction within the accuracy stated above.

\newpage{}

\begin{figure}[h]
\begin{raggedright}
\includegraphics[width=10cm]{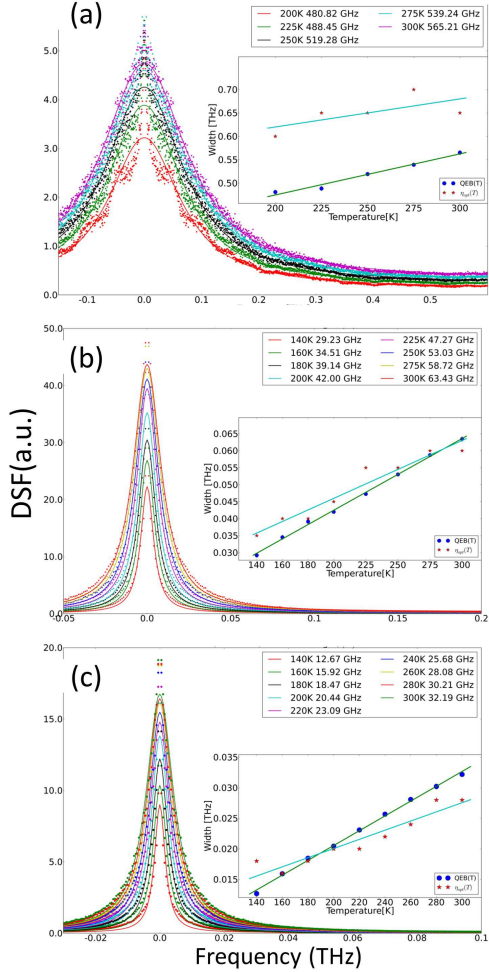}\\

\par\end{raggedright}

\protect\caption{\label{fig:QEB MD}From top to bottom, results of the $QEB_{MD}(T)$
fit to a single Lorentzian for 23, 100 and 200 amu adsorbate. The
inset shows the QEB widths extracted from MD data simulations at different
temperatures alongside the optimal friction values,  $\eta_{opt}$, used by the Langevin
simulation to fit the MD $\Gamma\left(\Delta K\right)$ curves .}
\end{figure}

Next, we applied the same procedure on the MD data in order to extract
effective friction values and check how they change with temperature.
Figure \ref{fig:QEB MD} shows the QEB peaks calculated for the different
temperatures and different adsorbate masses, and a Lorentzian peak
fit (full lines) to the QEB. First we note, that the Lorentzian fit
to the QEB peak is not quite as good as it was for the spectra calculated
by the Langevin simulation, mostly due to low frequency peaks related
to the surface vibrations and an incomplete subtraction of the QEP
peak (assumed to be a delta function at the diffraction condition).
Nevertheless, we see that the values extracted from the Lorentzian
width are quite close to the Langevin friction values which were obtained
by fitting the $\Gamma\left(\Delta K\right)$ curves ($\eta_{opt}$). The inset depicts the comparison between $\eta_{opt}$
and QEB widths (extracted from the MD simulation) for the different
temperatures and adosrbate masses. An obvious feature which can be
seen from these graphs is that for all three masses the QEB widths
extracted from the MD calculations increase as function of temperature,
more or less following the trend of $\eta_{opt}$. Consequently ,
we conclude that the need to increase $\eta_{opt}$ as function of
T when trying to reproduce the MD results with the Langevin simulation,
represents a property of the frictional coupling of the MD simulation,
which is then revealed in the comparison with the Langevin simulations.
In other words, the fact we had to increase $\eta_{opt}$ with temperature
in order to mimic the MD results with the Langevin code, does not
indicate a discrepancy between the two simulations.

\section{Summary and Conclusions}

We have compared two different numerical approaches for interpreting
adsorbate diffusion on a solid substrate, namely, MD and Langevin
simulations. A major difference between these two approaches is the
substitution of the dynamic substrate which is explicitly simulated
by the MD code, with a friction damping term and a stochastic force
in the Langevin simulation. Since this substitution can not accurately
account for correlations between the relative motion of the substrate
and adsorbate atoms which takes place in the MD, a certain discrepancy
in the simulated dynamics is anticipated. For example, a substrate
phonon creates a time dependent distortion of the potential energy
surface on which the adsorbate moves, hence one could expect that
the rate of single jumps and longer jumps would be affected by the
frequency and amplitude of the substrate vibrations. While it is obvious
that such correlations will take place to some degree, a quantitative
assessment of the discrepancies was missing in the literature, and
it was unclear whether they are sufficiently large to affect the interpretation
of realistic (noisy) experimental data. The observables we chose to
compare are the ISF and DSF correlation functions, focusing on the
width of the quasi-elastic peak, $\Gamma$, in particular. The dependence of the
quasi-elastic peak width on the momentum transfer and sample temperature
provides a sensitive measure of the motion rate and mechanism, and
is also accessible to helium scattering experiments\cite{alexandrowicz_helium_2007,jardine_studying_2009,jardine_helium-3_????}. 

The comparison we performed showed that for the particular systems
we simulated, the two simulations can produce very similar observables, using the friction parameter as the only free parameter
used to fit the two. Thus, within the conditions we simulated, correlation
effects do not seem to lead to any noticeable discrepancies between
the two simulations, and the Langevin simulations provides a good approach for simulating the surface dynsmics. 

We did notice that the optimal friction values
which we obtained from fitting the two simulations increased slightly
with temperature, an effect which was more significant for adsorbates
with a higher mass. One possible interpretation of this observation
is that the need to increase the friction parameter of the Langevin
simulation at higher temperatures is an indication of a discrepancy
between the two numerical approaches (i.e. we are compensating for
fundamental differences between the two simulations by adjusting the
fit parameter). Another interpretation is that the frictional coupling
rate is increasing with temperature in the MD simulation and that
the comparison with the Langevin simulation (which produces an optimal
friction parameter) is simply revealing this fact. We used the quasi-elastic
base width as a method to estimate the frictional coupling from the
dephasing rate of the motion within the adsorption site and extract
effective friction values from the MD data. Our results show that
the effective friction values extracted using this method, are in
close proximity to those used to fit the Langevin simulation. Furthermore,
the effective friction values also show an increase with temperature
supporting the second interpretation mentioned above, i.e the frictional
coupling increases with temperature in the MD simulation and the need to adjust the
friction parameter of the Langevin simulation to fit the two simulations
does not indicate a discrepancy between the two numerical approaches. 

In conclusion, when using the particular interaction models mentioned
above, adsorbate masses ranging from 23 to 200 amu, and temperatures
within the range of 140K to 300K, we do not observe significant differences
between the Langevin and MD simulations. Thus, even if differences
exist, they are subtle and should not affect the analysis of experimental
data with similar or larger noise levels. An explanation for this
lack of discrepancy, might be that the relatively fast time scales
which characterize the substrate motion lead to an averaging effect
which reduces the importance of explicit correlations and allows us
to treat the interaction as a sum of a static interaction (PES) and
a stochastic force with a good accuracy. It is also worth noting,
that in the past when applying Langevin simulations for data analysis,
it was assumed that the friction is independent of surface temperature.
While the particular trend we observed in the MD simulation reflects
our choice of model for simulating the substrate (harmonic potential)
and adsorbate (Morse potential) and is not directly related to other
systems and interaction models, it is worth remembering that the friction
might change as function of temperature also in other systems. If
this temperature dependence is not negligible and is not taken into
account, systematic errors might be produced when extracting physical
properties from the simulations, in particular the energy barrier
for diffusion deduced from Arrhenius graphs. Finally, we assume that
there will be other systems and conditions under which correlations
will produce noticeable effects, however, these will probably require
substantially different time scales (faster adsorbate motion or slower
substrate motions) and it should be interesting to study such systems
in the future.

\section{Acknowledgements}

The authors would like to thank Prof. Erio Tossati for valuable scientific
discussions. This work was supported by the Israeli Science Foundation
(Grant No. 2011185) and the European Research Council under the European
Union\textquoteright s seventh framework program (FP/2007-
2013)/ ERC Grant 307267.

\bibliographystyle{unsrt}


\end{document}